\documentclass[lettersize,journal]{IEEEtran}
\usepackage{amsmath}
\usepackage{amsfonts}
\usepackage{amssymb}
\usepackage[ruled,linesnumbered]{algorithm2e}
\usepackage{array}
\usepackage{booktabs}
\usepackage{bm}
\usepackage{mathtools}
\usepackage[caption=false,font=normalsize,labelfont=sf,textfont=sf]{subfig}
\usepackage{textcomp}
\usepackage{stfloats}
\usepackage{witharrows}
\usepackage{url}
\usepackage{verbatim}
\usepackage{graphicx}
\usepackage{cite}
\usepackage{enumerate}
\usepackage[inline]{enumitem}
\usepackage{setspace}
\usepackage{makecell}

\begin{document}

\title{Make Safe Decisions in Power System: Safe Reinforcement Learning Based Pre-decision Making for Voltage Stability Emergency Control}

\author{Congbo Bi,~\IEEEmembership{Student Member,~IEEE,} Lipeng Zhu, Di Liu,~\IEEEmembership{Member,~IEEE,} and Chao Lu,~\IEEEmembership{Senior Member,~IEEE} 
		\thanks{This work was supported in part by the China Southern Power Grid Research Project ZBKJXM20232029 and the National Natural Science Foundation of China under Grant 52207094.}
		\thanks{This work has been submitted to the IEEE for possible publication. Copyright may be transferred without notice, after which this version may no longer be accessible.}
		 \thanks{Congbo Bi is with the Department of Electrical Engineeering, Tsinghua University, Beijing, 100084 China (e-mail: thueea\_bcb@outlook.com).} 
		 \thanks{Lipeng Zhu is with the College of Electrical and Information Engineering, Hunan University, Hunan, 410082 China (e-mail:zhulpwhu@126.com).} 
		 \thanks{Di Liu is with the Department of Electrical Engineeering, Tsinghua University, Beijing, 100084 China (e-mail: kfliudi@163.com).}
		 \thanks{Chao Lu is with the Department of Electrical Engineeering, Tsinghua University, Beijing, 100084 China (e-mail: luchao@tsinghua.edu.cn).}
	}
	\markboth{}%
	{Shell \MakeLowercase{\textit{et al.}}: }
	
	\maketitle

\begin{abstract}
The high penetration of renewable energy and power electronic equipment bring significant challenges to the efficient construction of adaptive emergency control strategies against various presumed contingencies in today's power systems. Traditional model-based emergency control methods have difficulty in adapt well to various complicated operating conditions in practice. Fr emerging artificial intelligence-based approaches, i.e., reinforcement learning-enabled solutions, they are yet to provide solid safety assurances under strict constraints in practical power systems. To address these research gaps, this paper develops a safe reinforcement learning (SRL)-based pre-decision making framework against short-term voltage collapse. Our proposed framework employs neural networks for pre-decision formulation, security margin estimation, and corrective action implementation, without reliance on precise system parameters. Leveraging the gradient projection, we propose a security projecting correction algorithm that offers theoretical security assurances to amend risky actions. The applicability of the algorithm is further enhanced through the incorporation of active learning, which expedites the training process and improves security estimation accuracy. Extensive numerical tests on the New England 39-bus system and the realistic Guangdong Provincal Power Grid demonstrate the effectiveness of the proposed framework.
\end{abstract}


\setlength{\abovedisplayskip}{1.5pt} 
\setlength{\belowdisplayskip}{1.5pt}

\section{Introduction}\label{sec:1}
\IEEEPARstart{W}{ith} the increasing penetration of renewable energy sources and electronic equipment into modern power systems, the operating conditions of the power systems are more and more complex and variable, bringing unprecedented challenges to the efficient construction of adaptive emergency control strategies against various presumed contingencies \cite{yiImpactHighPenetration2019}. Conventionally, the widely-used approach to emergency control in practical power systems is implemented in the form of formulating formulating a set of control strategies in advance, and matching them with practical operational scenarios in real-time. This procedure involves the construction of a `presumed fault set’, necessitating numerous time-consuming simulations to derive a suitable table of emergency control strategies \cite{ZhangHuBeiHuangGangHuangShiDiQuAnQuanWenDingKongZhiXiTongFangAnSheJi2010}. Considering the complexity, it is a common practice to focus on a small number of representative scenarios for strategy formulation. As the structural and operational complexity of power systems increases, the above-mentioned traditional approaches is confronted with challenges of how to effectively manage the remendous presumed operational scenarios within a limited period of time \cite{zhaoDianlixitongtongyongankongceluezhengdingfangfadeyanjiu2015}. Moreover, the intricate dynamic characteristics of renewable energy sources and power electronic devices make it quite difficult to accurately model system dynamics, thereby undermining the reliability of the traditional approach to emergency control\cite{shairPowerSystemStability2021}. In this respect, it is imperative to swiftly identify high-risk operating statuses and formulate suitable emergency control measures.

Different from traditional approaches severely relying on detailed system modeling and simulations, deep reinforcement learning (DRL) methods provide a promising data-driven alternative to solve these issues. Specifically, DRL agent can learn from interactions with the environment of the system and update its policy without reliance on the knowledge about detailed system models and parameters \cite{sutton2018reinforcement}. There have been extensive studies on the formulation of DRL-based emergency control strategies \cite{liuJiyushenduqianghuaxuexidedianwangjinjikongzhicelueyanjiu2018, zhangDeepReinforcementLearning2021a, hossainGraphConvolutionalNetworkBased2021}, where emergency control is modeled as a Markov decision process (MDP) and a DRL agent learns optimal policy in a trial-and-error fashion \cite{chenReinforcementLearningSelective2022}. Nonetheless, most DRL methods lack explicit safety guarantees of the control strategies, which may limit their application in actual power systems \cite{dulac-arnoldChallengesRealworldReinforcement2021}. Moreover, the prevalent offline training procedure of RL agents requires the preparation of substantial learning samples, raising challenges in offline-learning efficiency and thus limiting its availability. Consequently, to enhance the applicability of DRL-based emergency control, it is imperative to introduce security constraints into DRL-based emergency control strategy formulation.

\IEEEpubidadjcol
For security constraints, most existing RL methods use a penalty term to discourage violations of the security constraints \cite{zhangDeepReinforcementLearning2021a, hossainGraphConvolutionalNetworkBased2021}, but this does not guarantee satisfying hard constraints. Additionally, some studies simplified the optimization problem by transforming constraints, e.g., constructing Lyapunov function \cite{shiStabilityConstrainedReinforcement2022} or using Lagrangian methods \cite{vuSafeReinforcementLearning2021a}. However, building a suitable Lyapunov function or solving the dual optimization problem is challenging in complex systems. To address this issue, safe reinforcement learning (SRL), incorporating security constraints into the RL framework, is proposed \cite{brunkeSafeLearningRobotics2022}. SRL has been applied to tackle some power system control and decision-making issues \cite{liResearchApplicationSafe2023, wangSafeOffPolicyDeep2020, gaoModelaugmentedSafeReinforcement2022}. However, existing studies still have some limitations. Firstly, their learning schemes are mainly designed for strategy formulation about steady-state scenarios like power system dispatch, which have difficulties in emergency control scenarios with much more complex dynamics. Secondly, these existing algorithms can't adequately address the nonlinear characteristics of transient constraints and feasible action space boundaries of emergency control. Therefore, adapting these methods to develop effective emergency control strategies would be challenging.

SRL addresses the feasibility of solutions; however, managing large datasets remains a pragmatic challenge in realistic applications. Active learning (AL) techniques mitigate this issue by efficiently leveraging extensive samples \cite{renSurveyDeepActive2021}. In contemporary data-driven power system emergency control, the critical samples not only refer to high-risk operating conditions with potentially hazardous control actions but also refer to conditions at the edge of the operating space. We predominantly address the former in this work. While numerous studies have explored the application of AL methodologies within the power system stability assessment \cite{malbasaVoltageStabilityPrediction2017,zhangPowerSystemTransient2021,wangPowerSystemTransient2022}, there exists a discernible void in the quantitative estimation of the security margin within emergency control decision-making processes. Integrating AL learning can enhance thesecurity margin estimators' learning efficiency and practical implementation in actual systems.

To address these challenges, a novel SRL-based emergency control strategy pre-formulation via security margin estimation and gradient projection is developed in this paper, where both the nonlinear transient stability constraints and strategies are approximated by neural networks (NNs). Considering that short-term voltage stability emergency control is one of the most challenging tasks in power systems, successful performance by the proposed method in this context provides compelling evidence for its effectiveness in other related tasks. Therefore, we choose short-term voltage stability emergency control as an illustrative example. This study conceptualizes short-term voltage stability emergency control as a state-constrained Markov decision process (SCMDP). The post-fault security margin of the pre-decision action is adjudicated via a NN-based estimator. For insecure actions, security margin gradient-based corrective measures are implemented until safe. Moreover, we augment the algorithm’s efficacy on both theoretical and practical fronts: theoretically, by furnishing proof of the algorithm’s security, and practically, by integrating AL to bolster the training efficiency of the security margin estimator amidst a voluminous dataset. The main contributions of our work are as follows:

\begin{enumerate*}[label = \arabic*. ,itemjoin=\\\hspace*{\parindent}]      
    \item We present an emergency control framework predicated on SRL, utilizing gradient projection to ensure short-term voltage stability in power systems. A theoretical performance proof accompanies this framework. Compared with its counterparts, our method has a more solid theoretical foundation and performs well in power systems with various scales. 
    \item We propose a security margin estimation scheme integrating state and action information in emergency decision-making scenarios for the first time. Compared to the fully connected net structure, the proposed structure can effectively capture the characteristics of risky operation points and has more robust adaptive capability against various operating scenarios.
    \item The proposed AL algorithm enhances the security margin estimator's training process by swiftly identifying critical operating points within complex power systems. This enhancement significantly boosts the security margin estimator's training efficiency. Moreover, its high practical value is underscored by its applicability to large-scale power systems for effectively capturing critical operating conditions.
\end{enumerate*}     

The remainder of the paper is structured as follows. Section \ref{sec:2} briefly introduces the basic knowledge of pre-decision making, SRL algorithm, and AL. Section \ref{sec:3} elaborates the proposed framework, including the AL-based security margin estimator and the gradient projection-based short-term voltage stability emergency control pre-decision making module. The New England 39-bus system and the Guangdong Provincal Power Grid (GPG) are utilized to test the performance of our framework in Section \ref{sec:4}. Section \ref{sec:5} concludes the paper.

\section{Preliminaries}\label{sec:2}
\subsection{Safe Reinforcement Learning (SRL)}
Reinforcement learning involves continuous interaction with the environment to ascertain the most effective strategy via trial and error. However, in some practical applications, including autonomous driving and power system operations, faulty policies can result in substantial economic repercussions or compromise safety \cite{brunkeSafeLearningRobotics2022}. Hence, the imperative to integrate safety considerations during the training or operational phases in such contexts has precipitated the emergence of SRL algorithms. Depending on the form of constraints, safe reinforcement learning can be categorized into two types: process-wise safe reinforcement learning and state-wise safe reinforcement learning. Initial endeavors of process-wise safe reinforcement learning were grounded on the constrained Markov decision processes (CMDP) framework, wherein constraints are articulated as either cumulative or episodic. However, real-world scenarios often involve transient and deterministic critical constraints, which, if violated, can lead to catastrophic task failure \cite{zhaoStatewiseSafeReinforcement2023}. Therefore, it is necessary to introduce more robust constraints in reinforcement learning. Unlike process-wise safe reinforcement learning, state-wise SRL is based on the SCMDP, where the safety specification is to satisfy a hard cost constraint at every step persistently. Similarly, the cost functions are denoted with $C_{1}, C_{2}, ...$, and the feasible stationary policy set $\bar{\Pi}_{C}$ for SCMDP is defined as:
    \begin{equation}
        \begin{aligned}
            \bar{\Pi}_{C} = \{\pi \in \Pi | \forall (s_{t},a_{t},s_{t+1}) \sim \tau, \\ 
            \forall i, C_{i}(s_{t},a_{t},s_{t+1}) \le \omega_{i}\}        
        \end{aligned}
        \label{eqn.2}
    \end{equation}
where $\tau \sim \pi$, $\omega_{i} \in \mathbb{R}$, and $\pi$ is the agent's policy. Compared to process-wise SRL, state-wise SRL has stronger constraint effects and is more suitable for scenarios with stricter state security demand.

The purpose of state-wise SRL is to find a policy that maximizes the cumulative gain $G = \sum_{t=0}^{\infty}{\gamma^{t}r_t}$ while satisfying the security constraints, which is mathematically expressed as
    \begin{equation}
        \mathop{\max}_{\theta} G(\pi_{\theta}), \quad 
 \mathrm{s.t.} \ \pi_{\theta} \in \bar{\Pi}_{C}
        \label{eqn.3}
    \end{equation}
where $r_t$ denotes the reward at $t$, $\gamma$ denotes the discount factor, and $\pi_{\theta}$ is the agent's parameterized policy.

\subsection{Pre-decision making and its SCMDP Modeling}\label{sec:2:2}
Pre-decision making plays a crucial role in preventing fault propagation and safeguarding power system security and stability. In this approach, emergency control strategies are pre-generated based on the system’s operating state and a predefined set of potential faults. When a fault occurs, the stability control apparatus consults these pre-established rules according to the fault type and the existing operating condition and then executes the necessary emergency control actions according to pre-determined criteria. Emergency control measures in power systems, such as load shedding, generator tripping, and modulating DC transmission line power, are essential for preventing system collapse and enhancing stability. In this context, we focus on load shedding as a representative example. Our proposed pre-decision making scheme aims to address short-term voltage instability, minimizing the impact of faults while keeping costs as low as possible. Specifically, we seek to reduce the amount of load shedding required to stabilize the system’s voltage after a fault. The mathematical formulation of the power system’s short-term voltage stability control pre-decision making process is delineated as follows:
    \begin{equation}
    \begin{aligned}
        \mathop{\arg \min}_{a_{t} \in A} C(s_{t},a_{t}) \\
        \text{s.t.} \quad f(s_{t+T})\le 0 \\
    \end{aligned}
    \label{eqn.1}
    \end{equation}
where $s_{t}$ represents the extant operating status of the system at time $t$ and $a_{t}$ signifies the emergency control action, e.g., load shedding amount at time $t$. The cumulative expense associated with $a_{t}$ is denoted as $C(s_{t},a_{t})$. The function $f(\cdot)$ encapsulates the transient stability constraints during the post-fault recovery phase. $f(s_{t+T})\le 0$ indicates the system remains stable following the fault and the emergency control actions.

The short-term voltage stability emergency control strategy pre-formulation can be modeled as a SCMDP. Critical elements in the SCMDP are defined as follows.
\subsubsection{State and Action} In practical power systems, the injected power space $[P,Q]_{1 \times 2n}$ is commonly utilized as the input information for emergency control pre-decision making \cite{ShiAssessmentofSystemProtectionStrategy2020}. Furthermore, the bus voltage $V$, the active power $P_{G}$, and reactive power generation $Q_{G}$ of generators are also incorporated as a critical piece of input information. Consequently, the state variables can be succinctly represented as $O(t)$:
    \begin{equation}
    O(t) = [P,Q,V,P_{G},Q_{G}]_{1\times(3n_{b}+2n_{g})}
    \label{eqn.3-1}
    \end{equation}
where $n_{b}$ and $n_{g}$ denote the amount of buses and generators in the system, respectively. The state information $s$ aligns with a specific point within the power system operating space. As per the principles of dynamic security region theory \cite{yuTheoryMethodPower2020}, the operating point is associated with a sequence of emergency control actions that fulfill the established constraints. Consequently, it is practical to devise an emergency control strategy that takes the power system's operating point as the primary input information and implements it via a real-time matching approach in realistic power systems.

The action space $A$ in emergency control pre-decision making refers to the load-shedding, adjusting the power on DC transmission lines, or generator-tripping. Since under-voltage load-shedding (UVLS) is a common control strategy to prevent voltage collapse, UVLS is considered as the emergency control measure taken in our proposed scheme: 
    \begin{equation}
    \bm{a}(t) = [\Delta L_{1}(t),...,\Delta L_{n_d}(t)]_{1\times n_d}
    \label{eqn.3-2}
    \end{equation}
where $\Delta L_{i}(t) $ denotes the proportion of load-shedding at the $i^{\rm th}$ load shedding control device, and $n_d$ denotes the device amount. Without loss of generality, this study's methodology and conclusions can be generalized to situations with more diverse control measures.
\subsubsection{Reward} The reward function of the short-term voltage stability emergency control is designed as follows:
    \begin{equation}
        r = 
        \left\{\begin{aligned}
                & -\xi, \min{v_{j}(t+10)} \leq 0.8\\
            &  - \alpha \sum\limits_{i=1}^{n_d} \Delta L_{i}(t) - \beta \sum\limits_{j=1}^{n_{b}} \Delta v_{j}^2,{\quad\rm otherwise }\\
        \end{aligned}
        \right.\label{eqn.3-3}
    \end{equation}
where $-\xi$ is a large negative number and $\alpha$,$\beta$ are the weight factors respectively. $\sum\Delta L_{i}(t)$ denotes the amount of load shedding, and $\sum\Delta v_{j}^2$ denotes the voltage deviation.
\subsubsection{State Constraint}The state constraint in short-term voltage stability control ensures the system is stable after a disturbance. Specifically, the post-fault state must be stable, as expressed in (\ref{eqn.1}).

Although the actual UVLS involves multiple rounds of load shedding, removing a sufficient number of loads simultaneously earlier can prevent the spread of faults more effectively \cite{larikCriticalReviewStateofart2019}. Consequently, our analysis focuses on single-round resection.

\subsection{Active Learning (AL)}
AL is a semi-supervised technique that identifies and queries the most informative examples, e.g., those with the highest uncertainty in classification tasks, during model training. AL operates on the premise that a model learns more effectively when it concentrates on ambiguous instances. Metrics such as entropy or probability distributions are employed to gauge and articulate the level of uncertainty. By emphasizing these uncertain points, AL enhances model accuracy using fewer labeled examples, thus optimizing labeling efforts and reducing costs. AL strategies can significantly reduce the amount of data needed for training by intelligently selecting the most informative samples, which saves resources and leads to more accurate and robust models.

\section{Proposed Methodology}\label{sec:3}
\subsection{Overall Framework}\label{sec:3:1}
\begin{figure}[tbp]
		\centerline{\setlength{\textfloatsep}{-5mm}\includegraphics{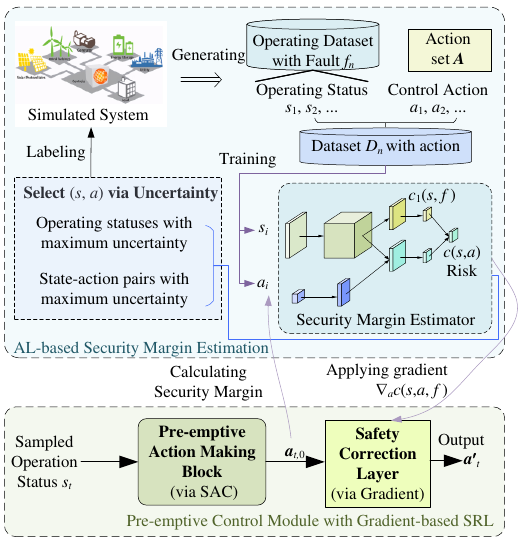}}
        \vspace{-0.3cm}
		\caption{Framework for proposed SRL-based pre-decision making scheme to against short-term voltage instability.}
        \vspace{-0.3cm}
		\label{fig1}
	\end{figure}

The overall framework of our proposed emergency control pre-decision making scheme via SRL is shown in Fig. \ref{fig1}, which includes two parts: the AL-based security margin estimator with a state-action joint network structure, and the projection-based pre-decision making module. In the proposed framework, the initial control actions generated by the DRL agent do not directly take effect. Instead, they pass through a security margin estimator constructed by a NN. Subsequently, these actions are adjusted based on the gradient of the estimated margin before being implemented.
\subsection{Active Learning-based Security Margin Estimation}\label{sec:3:3}
It is widely considered that short-term voltage instability often stems from inadequate local reactive power support \cite{dwivediLiteratureSurveyShortTerm2018}. In a receiving-end system, the post-fault stability is intricately linked to the pre-fault operating status. Additionally, the efficacy of emergency control measures implemented to mitigate fault progression is contingent upon the initial status. Consequently, estimating the system’s post-fault stability and the `margin' is achievable by checking the operational points and the corresponding emergency control measures \cite{chenPowerSystemTransient2023}. This subsection will detail the architecture of the envisioned security margin estimation module, encompassing the basic theory, the network structure, the pivotal samples selecting methodology, and the training methodology.

\subsubsection{Security Margin Estimation via DSR}
        \begin{figure}[tbp]
	\centerline{\setlength{\textfloatsep}{-5mm}\includegraphics{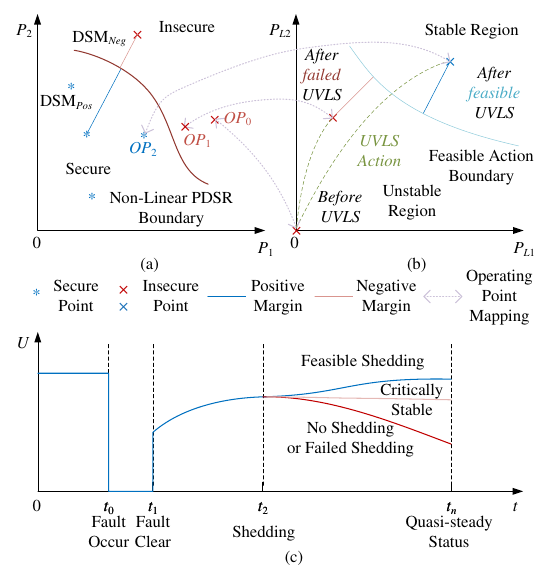}}
        \vspace{-0.3cm}
		\caption{Power system security margin incorporating both system state information and emergency control actions.}
        \vspace{-0.3cm}
		\label{fig1-1}
	\end{figure}
The theory of security regions \cite{yuTheoryMethodPower2020} maintains that the power system’s security region $A_{DSR}$ is well-defined within the context of injected power space for a given network topology, system component parameters, and the location of pre-determined faults. Without emergency control measures, the operating point's dynamic security margin (DSM) is quantified as the distance from the operating point to the dynamic security region boundary. Since the security region is delineated by hyperplanes, the computation of the DSM bifurcates into two processes: firstly, the delineation of the security region’s boundary $\partial A_{DSR}$, and secondly, solving for the minimum distance from the operational point to points on the designated boundary, which is shown in Fig.~\ref{fig1-1}(a). The definition of DSM is as follows:
    \begin{equation}
    \mathrm{DSM} = \min \mathrm{dis}(P, P_{C}), \quad \mathrm{s.t.} \ P_{C} \in \partial A_{DSR}
        \label{eqn.3-4}
    \end{equation}
where $P$ is the operating point in injected power space and $P_{C}$ is a point on the boundary. The term `dis' refers to Euclidean distance. This representation transforms the solution of the distance from the point to $\partial A_{DSR}$ into the solution of the minimum distance from the point to a point on $\partial A_{DSR}$. The detailed computational procedures of DSM are thoroughly outlined in \cite{chenPowerSystemTransient2023}. 

To enhance reader comprehension, we illustrate the concept using a system equipped with induction motors as an example. Specifically, we apply a similar mathematical formulation to address short-term DSM when considering emergency control actions. In this system, voltage instability is closely linked to the risk of motor stalling \cite{zhangLoadStabilityIndex2019}. Throughout the fault recovery stage, the system’s slip rate and the absorbed reactive power are approximately adhered to the subsequent equations:
    \begin{equation}
    \begin{aligned}
    2H_{i}\frac{ds}{dt}& = T_{m_{i}}-\frac{r_{R_{1}}v_{i}^2/s}{r_s+(r_{R_{1}}/s)^2+(x_s+x_{R_{1}})^2} \\
    q_i& = \frac{v_i^2}{x_m}+v_i^2\frac{x_s+x_{R_{1}}}{r_s+(r_{R_{1}}/s)^2+(x_s+x_{R_{1}})^2} \\
    \end{aligned}
    \label{eqn.3-5}
    \end{equation}
        \begin{figure}[tbp]
	\centerline{\setlength{\textfloatsep}{-5mm}\includegraphics{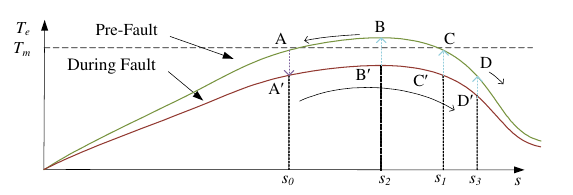}}
        \vspace{-0.3cm}
		\caption{Changes in slip during the fault process.}
        \vspace{-0.3cm}
		\label{fig1-2}
	\end{figure}
where $r_{R_{1}}$, $r_{s}$, $x_{R_{1}}$, $x_{s}$ and $x_{m}$ are the induction motor parameters. $T_{m_{i}}$ is the torque. $v_{i}$ is the voltage amplitude of bus $i$, and $q_{i}$ is the reactive power absorbed by the motor. From the given formula, the relationship between motor slip and electromagnetic torque $T_e$ can be determined. During the entire fault process, the slip of the motor changes as shown in Fig.~\ref{fig1-2}, where the slip ratio at the moment of the fault occurrence is $s_0$, and the slip ratio of the extreme point is $s_2$. When a fault occurs, the slip ratio will remain $s_0$, and the status point moves from A to A$'$. Because $T_e$ is smaller than $T_{m_{i}}$ at the moment, $s$ will increase. The slip ratio will remain unchanged when the fault is cleared, while the voltage will try to increase and revert to its value before the fault, altering the status point. During the recovery process, if the load shedding is not applied or the load shedding amount is insufficient, the slip ratio $s$ will continue to escalate, surpassing $s_2$; this leads to motor stalling and a failed UVLS scenario, depicted as ‘failed UVLS’ in Fig.~\ref{fig1-1} (b) (corresponding to point $OP_1$ in Fig.~\ref{fig1-1} (a)). Conversely, if the load shedding is sufficient, the motor slip will consistently remain below $s_2$ throughout the recovery, ensuring stability. This scenario is illustrated as `feasible UVLS’ in Fig.~\ref{fig1-1} (b) (corresponding to point $OP_2$ in Fig.~\ref{fig1-1} (a)). In the analysis outlined above, a critical load shedding threshold exists, which is the minimum amount of load that must be shed to ensure the system’s stability. This pivotal value is demonstrated as the `Feasible Action Boundary' in Fig.~\ref{fig1-1} (c). The dynamic action-jointed security margin (DASM) is defined as follows:
    \begin{equation}
    \mathrm{DASM} = \mathrm{dis}(P,\partial A_{DSR}) + \mathrm{dis}(P_{LS},\partial A_{LS,P}) 
        \label{eqn.3-6}
    \end{equation}
where $\partial A_{LS,P}$ is the `Feasible Action Boundary' of emergency control action at operating point $P$, and $P_{LS}$ is the load shedding amount. The latter term shows the impact of load shedding on the security margin, and the subscript $P$ denotes that the calculation of the threshold above depends on the specific operating point. Within the injected power space and the UVLS space, we can further simplify the solution to the formula as follows:
    \begin{equation}
    \begin{aligned}
        \mathrm{DASM} & = \min \mathrm{dis}([P, P_{LS}], [P_C, P_{LS,P}]), \\ 
        \mathrm{s.t.} & \ P_{C} \in \partial A_{DSR}\ \mathrm{and} \ P_{LS,P} \in \partial A_{LS,P}
    \end{aligned}
        \label{eqn.3-7}
    \end{equation}
Similarly, we can address the problem via gradient optimization. Consequently, this approach leads to an integrated representation of the power system’s security margin, which encompasses both the system’s state information and the emergency control measures.

\subsubsection{State-action-joint Structure for Margin Estimation}
        \begin{figure}[tbp]
		\centerline{\setlength{\textfloatsep}{-5mm}\includegraphics{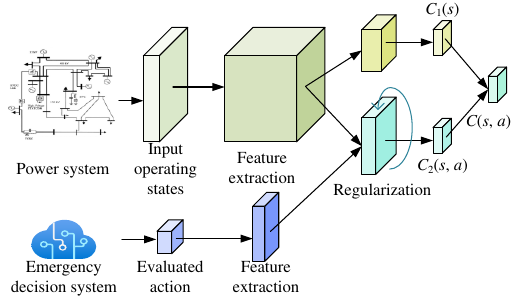}}
        \vspace{-0.3cm}
		\caption{Proposed state-action-joint security margin estimation network.}
        \vspace{-0.3cm}
		\label{fig2}
	\end{figure}
 
Despite introducing the security margin representation incorporating action and operating points, resolving the DASM with nonlinear attributes remains challenging. The employment of NNs, with their robust nonlinear feature extraction capabilities, presents a viable solution. However, the precise estimation of the DASM necessitates effectively extracting the power system’s operating point characteristics and their integration with emergency control actions. Drawing inspiration from the dueling DQN \cite{wangDuelingNetworkArchitectures2016}, our study reevaluates the challenges associated with short-term voltage security margin estimation in power systems. We focus on the discrepant impacts of the power system’s static operating point and the emergency control actions on post-fault stability. Specifically, we propose a joint state-action security margin estimation network, as shown in Fig.~\ref{fig2}. The proposed innovative \emph{dueling} framework delineates the state-wise security value function $C_{1}(s)$ and the advantage function $C_{2}(s,a)$ from the predicted overarching security margin $C$. The state-wise security value function $C_{1}(s)$ is exclusively tasked with forecasting the security margin at the present operating point, while the advantage function $C_{2}(s,a)$ is dedicated to evaluating the influence of each emergency control action on the security margin outcomes at that operating point:
    \begin{equation}
    C(s,a) = C_{1}(s) - C_{2}(s,a)
    \label{eqn.4}
    \end{equation} 
    
Given the absence of constraints on the state-wise security value function and the advantage function, (\ref{eqn.4}) is unidentifiable in the sense that $C_{1}$ and $C_{2}$ cannot be recovered via given $C$. Hence, the learned features may diverge from anticipated outcomes. To address this, (\ref{eqn.4}) is reformulated as follows:
    \begin{equation}
    C(s,a) = 
    \left\{\begin{aligned}
    & C_{1}(s) - (C_{2}(s,a) - \mathop{\min}_{a_{i} \in A} C_{2}(s,a_{i})), \rm{discrete} \\
    & C_{1}(s) - (C_{2}(s,a) - C_{2}(s,a_{zero})), \rm{continuous}
    \end{aligned}
    \right.\label{eqn.4-1}
    \end{equation} 
where the terms `discrete' and `continuous' refer to whether the action space is discrete or continuous. $A$ is the action space and $\mathop{\min}_{a_{i} \in A} C_{2}(s,a_{i})$ denotes the minimal advantage value in $C_{2}(s,\cdot)$ with discrete action space. Similarly, $a_{\text{zero}}$ epitomizes the advantage function under unregulated situations with continuous action space. The procedures above serve not only to regularize the feature extraction but also to fortify the training process.
 
Employing the decomposition mentioned above enables a more effective capture of the distinctive features of risky operating points within the power system, surpassing the conventional network architecture and refining estimation accuracy. Moreover, the proposed architecture’s capacity to independently investigate the attributes of operating states and actions ensures that the computation of the state-wise margin function and the advantage function remains robust against interference  upon changes in the operating scenarios. This attribute endows the proposed method with superior generalization capabilities across complex scenarios.

\subsubsection{Critical Operating Samples Searching and Training via Uncertainty}
Given the vast array of potential operating points in intricate power systems, one crux of enhancing training efficiency lies in selecting samples pivotal for training. In training the security margin estimator for obtaining the DSR boundary, not every sample contributes equally. Samples proximate to the classification boundary are instrumental in developing precise security boundaries and margins. Consequently, this research proposes a search scheme for pivotal operational samples grounded in AL methodology. The quintessence of AL involves commencing with a limited set of initial calibration samples to derive a preliminary evaluation model. Subsequently, employing a defined sample search strategy, a cohort of the most informative samples is selected for calibration. Specifically, a sample’s proximity to the DSR boundary directly correlates to its utility in obtaining an accurate boundary. The closer the operating point outcome of an evaluated sample is to the boundary, the more significant it is deemed for acquiring a precise estimator. Taking the threshold value $\epsilon$ for illustrative purposes, suppose the DASM of a single sample $(s,a)$ is denoted by $d(s,a)$. In this context, the uncertainty $U_{r}$ of the sample is defined as follows:
    \begin{equation}
    U_{r}(s,a) = \| d(s,a) \|^{-1}
    \label{eqn.5}
    \end{equation} 
Based on the state-action joint network structure and the definition of sample uncertainty, we introduce a robust training framework tailored for the proposed action-jointed security margin estimator leveraging AL to enhance its efficacy. The procedural steps are as follows:

\begin{enumerate*}[label = \arabic*. ,itemjoin=\\\hspace*{\parindent}] 
    \item \emph{Initial Sample Selection}: Randomly select $N$ samples to form the initial calibration data set for training the estimator.
    \item \emph{Initial Labeling}: Employ a simulation program to label the initial dataset with `stable' and `unstable' for training.
    \item \emph{Model Training}: Utilize the latest training sample set to train the DSR boundary and security margin estimator.
    \item \emph{Uncertainty Measurement}: Determine the uncertainty of unlabeled samples via the updated security margin estimator according to (\ref{eqn.5}). Identify and select the top $M$ unlabeled samples with the highest uncertainty for further processing.
    \item \emph{Simulation and Labeling}: Send the selected samples with the highest uncertainty to the simulator for labeling.
    \item \emph{Iterative Process}: Repeat steps 3 through 5, refining the calibration, training, and sample selecting procedures until the classification accuracy attains the preset threshold.
\end{enumerate*}     
 
The proposed methodology leverages an AL paradigm to expedite the learning process of the parameterized DASM estimator. Through iterative refinement and continual model updates, it demonstrates the capability to assimilate the intricacies of security assessment with a limited dataset. This approach not only streamlines the sample calibration process and improves the efficiency of constructing the security assessment model but also enhances the overall model training and updating speed, thereby contributing to the applicability of the proposed security margin estimator. Concurrently, the selection and utilization of critical samples enhance the module’s robustness amidst intricate scenarios.
\subsection{Gradient Projection-based Decision Making Module}\label{sec:3:4}
The quandary of fulfilling nonlinear safety constraints while concurrently ensuring performance parity remains a critical issue in SRL. A novel algorithm called unrolling safety layer (USL) is raised in \cite{zhangEvaluatingModelFreeReinforcement2023}, proffering a methodical enforcement of stringent constraints via the deep unrolling architecture and gradient projection. Inspired by USL, a two-stage short-term voltage stability emergency control policy pre-formulation and correction scheme is developed. The details of this scheme are shown in Fig.~\ref{fig4} and Fig.~\ref{fig4-1}.
        \begin{figure}[tbp]
		\centerline{\setlength{\textfloatsep}{-5mm}\includegraphics{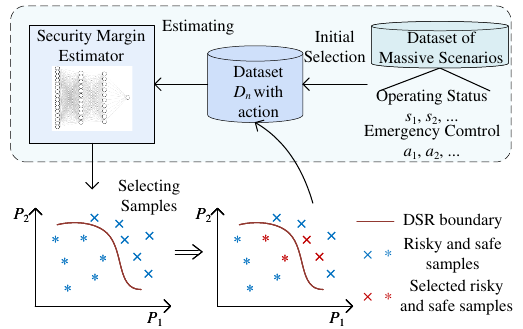}}
        \vspace{-0.3cm}
		\caption{Active learning-based security margin estimator training process.}
        \vspace{-0.3cm}
		\label{fig3}
	\end{figure}
        \begin{figure}[tbp]
		\centerline{\setlength{\textfloatsep}{-5mm}\includegraphics{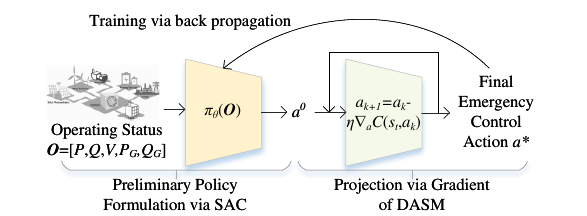}}
        \vspace{-0.3cm}
		\caption{Proposed gradient projection-based action correction.}
        \vspace{-0.3cm}
		\label{fig4}
	\end{figure}
        \begin{figure}[tbp]
        \vspace{-0.3cm}
		\centerline{\setlength{\textfloatsep}{-5mm}\includegraphics{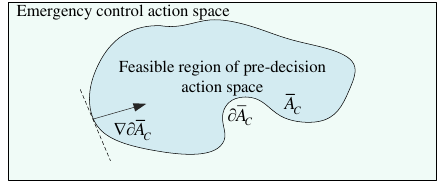}}
        \vspace{-0.3cm}
		\caption{Action correction via gradient.}
		\label{fig4-1}
	\end{figure} 
\subsubsection{Initial Emergency Control Policy Formulation}
Considering the multitude of operating points typically present in real-world power systems, it is imperative to to distill pivotal initial strategies from extensive operational datasets expediently. To this end, the formulation of the initial policy is predicated on the Soft Actor-Critic (SAC) algorithm, which is renowned for its efficacy in balancing exploration and exploitation in complex, high-dimensional spaces. The input of the proposed emergency control policy formulation network is the operating status $O(t) = [P,Q,V,P_{G},Q_{G}]$, and the output of the policy network is the uncorrected emergency control action $a^{0}$, i.e., the output action $a^{0}(t)$ is specified by a parameterized NN $\pi_{\theta}(O_t)$. The details of SAC can be found in \cite{haarnojaSoftActorCriticOffPolicy2018}. 

The reward mechanisms outlined in \ref{sec:2:2} are instrumental in guiding the decision-making agent away from hazardous strategies throughout this stage. Despite these measures, there remains a possibility of constraint violations occurring, which is attributed to the inherent challenges of RL agents when dealing with intricate nonlinear constraints, distributional shifts, etc. To tackle these challenges, we improve the existing framework. The preliminary decision network does not directly furnish the optimal action that meets all constraints. Instead, it provides an initial action for the subsequent correction stage, which may be sub-optimal but serves as a foundation for further refinement.

\subsubsection{Policy correction via Gradient Projection}\label{sec3:2:2}
Built on the premise that the security margin estimator acquires a relatively precise understanding of assessment knowledge, the action $a^{0}$ is adopted as the initial solution for the ensuing constrained optimization problem:
    \begin{equation}
        \begin{aligned}
            \mathop{\arg \max}_{a_{t} \in A} r(O_{t}, a_{t}) \\
            \text{s.t.} \quad D_{\theta}(O_{t}, a_{t}) \ge \epsilon\ \\
        \end{aligned}
        \label{eqn.6}
    \end{equation} 
where $r$ is the reward, $\epsilon$ is a positive margin value indicating secure status and $D_{\theta}(O_{t}, a_{t})$ is the estimated security margin (distance) at operating status $O_t$ with emergency control action $a_t$. The optimization problem in (\ref{eqn.6}) has nonlinear constraints and objective function. Considering the nonlinear characteristics of the problem, a direct linearization and projection could result in significant errors. We employ gradient descent to derive emergency control actions with an improved security margin to mitigate this. Specifically, the gradient of the parameterized security margin function, $\nabla_{a} D_{\theta}(s_{t}, a_{t})$, indicates the different impact of these control action components on the parameterized security margin, as follows:
    \begin{equation}
    \nabla_{a} D_{\theta}(s_{t}, a_{t}) = \{\frac{\partial D_{\theta}}{a(1)},\frac{\partial D_{\theta}}{a(2)},...,\frac{\partial D_{\theta}}{a(n)} \}|_{a = a_{t}}
    \label{eqn.7}
    \end{equation}
where $a(i)$ denotes the $i^{th}$ component of $a$. The above gradient information provides a reference for action correction. Subsequently, utilizing the gradient, we project the preliminary action into the feasible action space’s interior until the evaluated security margin $D_{\theta}(s_{t}, a_k)$ is no lower than a preset threshold $\epsilon$:
    \begin{equation}
    a_{k+1} = a_{k} + \lambda \nabla_{a} D_{\theta}(s_{t}, a_k)
    \label{eqn.8}
    \end{equation} 
where $\lambda$ is the step size. Corrective action aims to enhance the DASM, ensuring that the pre-decision action remains inherently more secure. The above gradient-based projections ultimately project the initial action into the feasible action space, as Fig.~\ref{fig4-1} shows. These operations transform risky actions into safer alternatives, and, finally, into a feasible emergency control action space. 

Furthermore, in this study, we propose a feasible update step size with no performance degradation via the Lipschitz continuous theory:
    \begin{equation}
    \lambda = \frac{D_{\theta}(s,a)-\epsilon}{2n_{a}L^2}
    \label{eqn.9}
    \end{equation} 
where $L$ is the Lipschitz constant of the network and $n$ is the dimension of pre-decision action. This result provides a solid theoretical foundation for our approach.

\section{Case Study}\label{sec:4}
The proposed scheme is tested on the IEEE 39-bus system and the GPG system in China Southern Power Grid (CSG) to illustrate the validity of the proposed methodology. All the simulations and tests are performed in a workstation equipped with an AMD Ryzen 9 5900X 12-Core Processor, 64GB RAM and an NVIDIA RTX 3090 GPU. The simulated system is built on Dynamic Simulation Program (DSP), an AC-DC hybrid power system calculation and analysis software developed by CSG Electric Power Research Institute (CSG EPRI), while the proposed SRL-based scheme is built on PyTorch 1.12.0.

\subsection{IEEE 39-bus System Case}
\begin{figure}[tbp]
		\centerline{\setlength{\textfloatsep}{-5mm}\includegraphics{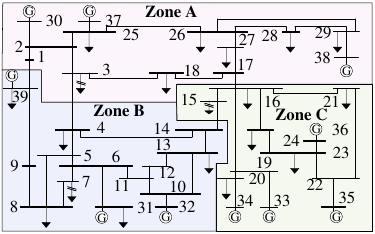}}
        \vspace{-0.3cm}
		\caption{Single line diagram of IEEE 39-bus system.}
        \vspace{-0.3cm}
		\label{fig5}
	\end{figure}

We first verified the performance of the proposed scheme via the IEEE 39-bus system, a simple approximation of the New-England Power System. The main structure of the IEEE 39-bus system is illustrated in Fig.~\ref{fig5}.
\subsubsection{Simulation Setup and Case Generation} 
Considering the operating point from \cite{athayPracticalMethodDirect1979} as the baseline condition, a variety of operating conditions and contingencies were generated. Specifically, the operating levels of the three zones varied from 60\% to 100\%, and a typical dynamic load proportion varied from 50\% to 60\%. In this study, we focus on faults occurring on lines across zones [F5, F19, F21] and faults in heavy-load areas [F7-18], and the fault clearing time was set to 0.18s. The proposed emergency control agents are specifically designed for each fault in the predefined fault set.
        \begin{figure}[tbp]
		\centerline{\setlength{\textfloatsep}{-5mm}\includegraphics{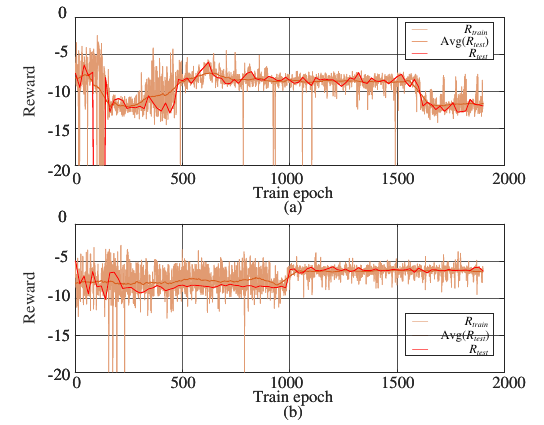}}
        \vspace{-0.3cm}
		\caption{Reward during the training process. (a) Reward with safety correction module. (b) Reward without safety correction module.}
        \vspace{-0.3cm}
		\label{fig5-1}
	\end{figure} 
\subsubsection{Validation of the SRL-based Decision Making Framework} 
Taking Fault F5 as an example, we tested the performance of the proposed scheme together with the one without correction. Fig.~\ref{fig5-1}(a) shows the reward in the training of the SRL-based emergency control pre-decision making module with $\epsilon=0.1$. Fig.~\ref{fig5-1}(b) shows the same training result without security correction. In the initial training phase, the proposed scheme outperforms the agent without safety correction because the proposed scheme integrates knowledge about feasible action space via DSM estimation, enabling efficient correction of risky actions. Despite initial transgressions of constraints in certain scenarios, the agent with security estimating faculties swiftly integrated knowledge about these incidents. This culminates in an enhanced proficiency of the proposed emergency control pre-decision making agent in navigating such predicaments. In the subsequent stage, the proposed scheme, bolstered by gradient projection, circumvented further violations and garnered a superior test reward relative to its counterpart devoid of action correction. The safety correction helps the agent stay within the safe range of actions, making the decision not only safer but also more stable and robust.
 
To validate the security of our proposed scheme, we compared its performance with that of the trained agent without action correction. This evaluation encompassed 100 scenarios unseen in training, including a random selection from a broader operating space and boundaries. These scenarios encompassed load deviations and dynamic load fluctuations to rigorously test the method’s dependability. The results, delineated in Table~\ref{table_1}, demonstrate that our proposed scheme had significantly fewer violations. In contrast, strategies employed by the agent without security correction failed to satisfy security constraints in a considerable proportion of scenarios, resulting catastrophic voltage collapse within the system. Moreover, adopting safe projection constraints guarantees that our strategy secures the system while necessitating only a minimal amount of additional load shedding, further substantiating the validity of our proposed scheme. Note that limited occurrences of violations occurred due to the security estimator's failure to capture some extremely infrequent cases, where our ongoing research will address this issue.
        \setlength{\tabcolsep}{3pt}	
        \vspace{-0.3cm}
        \begin{table}[htb]
        \small
		\caption{Performance Validation in Extreme Scenarios}
		\label{table_1}
		\centering
        \vspace{-0.2cm} 
        \begin{tabular}{cccccc}
			\toprule
			Method & Avg($R$) & Time/ms & Violation & Avg($\Delta v^2$) & Avg($P_{LS}$)\\
			\midrule
			Proposed & -12.83 & 1.33 & 6 & 0.0025 & 0.4349\\
			  SAC & -30.61 & 1.24 & 26 & 0.1152 & 0.3731\\ 
			\bottomrule
		\end{tabular}
        \vspace{-0.2cm} 
	\end{table} 
 
In addition, we compare our proposed safe decision-making method with other typical approaches, including the typical decentralized multi-round approach in \cite{larikCriticalReviewStateofart2019}, the traversal method, the Lyapunov-based safe policy optimization method \cite{chowLyapunovbasedSafePolicy2019} and Conservative Safety Critics (CSC) \cite{bharadhwajConservativeSafetyCritics2021}. The typical multi-round LS approach adheres to the following guidelines: If the voltage falls below a specified threshold within a certain duration, a predetermined amount of load is shed. The traversal method computes load shedding amounts at each bus using a fixed 20\% step. We maintain consistent hyperparameters for the latter two methods based on our proposed scheme. Table~\ref{table_4} summarizes their pre-decision making performances, where $P_{LS}$ denotes load shedding amount. Due to time constraints, we did not evaluate the traversal method on all available data. Our results demonstrate that our proposed method outperforms the other two approaches. This superiority can be attributed to the proposed method’s integration of nonlinear constraints and policy functions, which ensures flexibility and adaptability to complex fault scenarios. In contrast, the typical multi-round LS approach neglects the impact of system dynamic characteristics on voltage recovery during subsequent periods. In certain instances, untimely load shedding led to voltage collapse after a duration. The CSC suffered from random strategy selection limiting its effectiveness in utilizing power system safety constraints and control actions, while Lyapunov linearization lacked suitability for nonlinear scenarios with high load over-shedding.
        \setlength{\tabcolsep}{3pt}	
        \begin{table}[htb]
        \small
		\caption{Performance of Different UVLS Approaches}
		\label{table_4}
		\centering
		\begin{tabular}{ccccc}
			\toprule
			Method & Avg($R$) & Time/s & Violation & Avg($P_{LS}$) \\
			\midrule
			Proposed & -4.671 &  0.00133 & 0 & 0.2009\\
                Typical LS & -20.04 & - & 38 & 0.3120\\
			Traversal & $\mathrm{-}$ & $>$1e2 & 0 & $\mathrm{-}$ \\ 
			Lyapunov PG & -6.154 & 0.00139 & 0 & 0.3231 \\
                CSC & -10.32 & 0.00329 & 0 & 0.8794 \\
			\bottomrule
		\end{tabular}
        \vspace{-0.2cm}
	\end{table}
\subsubsection{Validation of the Proposed Security Margin Estimator} 
We evaluate the effectiveness of our proposed security classifier and margin estimator structure by comparing it with traditional alternatives, including fully connected network (FCN), and Long Short-Term Memory (LSTM). Our testing dataset comprises 3500 samples, representing random operating points and emergency control actions. For all networks, we set the learning rate to 1e-4, use 3 layers, and allocate 256 neurons per layer unless specific considerations dictate otherwise.

The performance for different network structures is summarized in Table \ref{table_2}, covering learning efficiency, accuracy, and specificity. Notably, the proposed network structure outperforms FCN across all evaluated metrics. While the proposed network may not match the performance of LSTM in certain respects, it’s essential to recognize that LSTM demands significantly more training time due to its complex architecture. Additionally, the risk categorization employed by the LSTM model tends to be overly binary, leading to potential misclassification of pivotal states. This issue becomes particularly pronounced under extreme operational conditions. In summary, our proposed security classifier effectively identifies risky operating points and control actions, aiding early warning and subsequent corrective measures.
	\setlength{\tabcolsep}{3pt}	
        \vspace{-0.3cm}
        \begin{table}[htb]
        \small
		\caption{Performance Comparison of Different Network Structures for Security Assessment and Margin Estimation}
		\label{table_2}
		\centering
        \vspace{-0.2cm}
		\begin{tabular}{ccccc}
			\toprule
			Method & Acc. & Spe. & Train Time/s \\
			\midrule
			Proposed & 0.988 &  0.978 & 27.48 \\
			FCN & 0.963 & 0.986 & 20.73 \\ 
			LSTM & 0.978 & 0.908 & 91.53 \\
			\bottomrule
		\end{tabular}
        \vspace{-0.2cm} 
	\end{table}
 
\subsubsection{Validation of the Proposed Active Learning Scheme} 
\begin{figure}[tbp]
		\centerline{\setlength{\textfloatsep}{-5mm}\includegraphics{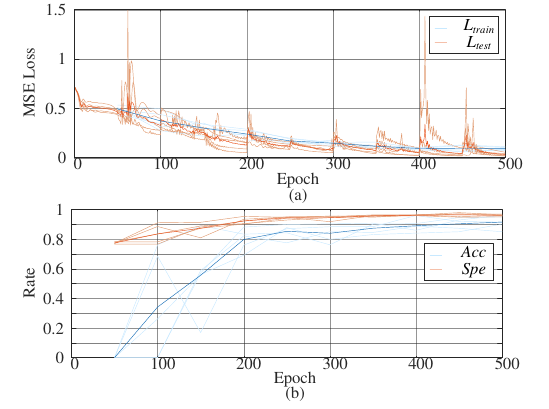}}
        \vspace{-0.3cm}
		\caption{Learning curves of security  via active learning. (a) Train and test losses (MSE). (b)Test accuracy.}
        \vspace{-0.3cm}
		\label{fig6}
	\end{figure}

\begin{figure}[tbp]
		\centerline{\setlength{\textfloatsep}{-5mm}\includegraphics{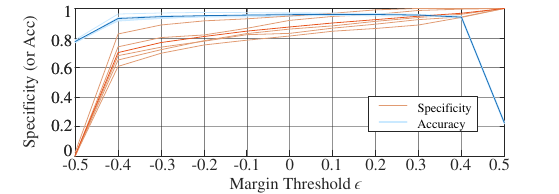}}
        \vspace{-0.3cm}
		\caption{Change in specificity and accuracy as $\epsilon$ changes.}
        \vspace{-0.3cm}
		\label{fig7}
	\end{figure}

In this subsection, we validate the proposed AL-based training efficiency enhancement scheme. Following the earlier settings, we generated 3500 simulation cases using time-domain numerical simulation. With a threshold of $\epsilon=0.5$ to distinguish between stable and unstable states, and 50 iterations for each selected set with 128 samples, we trained the proposed security margin estimator using 5-fold cross-validation (as depicted in  Fig.\ref{fig6} and Fig.\ref{fig7}). During training, both the training error and testing error decreased gradually, leading to an accuracy exceeding 95\% on the test set. 

Furthermore, we compared the proposed scheme with an alternative approach trained using all cases. The results, summarized in Table III, highlight the learning performance achieved by our method within 300 training epochs (equivalent to 384 cases, considering a batch size of 128). In contrast, the alternative approach necessitated over 5,000 seconds for dataset preparation and exhibited much longer training time (7.142s vs 50.76s). These findings underscore the effectiveness of our proposed method, which can effectively adapt to changing operating scenarios and topologies.
 
	\setlength{\tabcolsep}{3pt}	
        \vspace{-0.3cm} 
        \begin{table}[htb]
        \small
		\caption{Efficiency Verification of Proposed Active Learning-based Security Assessment and Margin Estimation Approach}
		\label{table_3}
		\centering
        \vspace{-0.2cm} 
		\begin{tabular}{ccccc}
			\toprule
			Method & Acc. & Spe. & $t_{\mathrm{total}}$/s & $t_{\mathrm{total}}$/s\\
			\midrule
			Proposed & 0.9602 &  0.8505 & 974 & 7.142\\
			Normal & 0.9786 & 0.9112 & 5850 & 50.76\\ 
			\bottomrule
		\end{tabular}
        \vspace{-0.2cm} 
	\end{table}

\subsection{GPG System Case}
\begin{figure}[tbp]
		\centerline{\setlength{\textfloatsep}{-5mm}\includegraphics{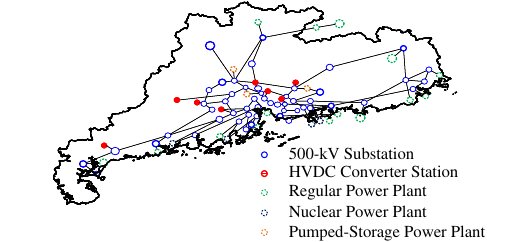}}
        \vspace{-0.3cm}
		\caption{Single line diagram of GPG system from realistic CSG.}
        \vspace{-0.3cm}
		\label{fig11}
	\end{figure}

To further validate the scalability and applicability of the proposed method in real power grids with complex structures and operating conditions, extended study in the GPG system is considered in this subsection. The GPG system case is simplified from one realistic provincial power grid of CSG, which is relatively susceptible to short-term voltage stabilization problems due to the concentration of numerous industrial areas with high electrical loads. Consequently, we selected the GPG system to investigate the proposed pre-decision making scheme. The structure of the GPG system is depicted in Fig.~\ref{fig11}, including all 500-kV buses and large power plants.
\subsubsection{Simulation Setup and Case Generation}
The overall load level of the GPG was set between $[0.6,1,1]$. Additionally, to account for potential load level fluctuations, we introduce supplementary disturbances at buses in heavy load areas to simulate the complex operating scenarios encountered in the realistic power systems. Furthermore, we intentionally set faults in the heavy load area to emphasize the voltage stability challenges faced by the system. Other settings are similar to the IEEE 39-bus system. Under these settings, we generate 5,000 samples, with 4,000 samples allocated for training and 1,000 samples reserved for testing.

\subsubsection{Overall Performance Validation and Comparison} 
\begin{figure}[tbp]
		\centerline{\setlength{\textfloatsep}{-5mm}\includegraphics{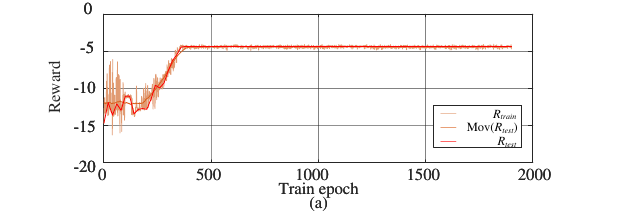}}
        \vspace{-0.3cm}
		\caption{Training process for GPG system.}
        \vspace{-0.3cm}
		\label{fig12}
	\end{figure}

The training and test results are depicted in Fig.~\ref{fig12} and Table \ref{table_5}. Considering the practical applicability of the proposed emergency control pre-decision-making scheme in complex realistic power systems, we conducted specific verification of its computational efficiency within the GPG system. Table \ref{table_5} also illustrates that the proposed method maintains robust performance in complex systems with zero violations. Simultaneously, it demonstrates greater efficiency compared to algorithms that seek the optimal reduction amount via exhaustive search. Despite the limited selection of test samples due to time constraints, these results corroborate its efficacy. 
 
	\setlength{\tabcolsep}{3pt}	
        \vspace{-0.3cm}
        \begin{table}[htb]
        \small
		\caption{Performance Validation in Extreme Scenarios for GPG System}
		\label{table_5}
		\centering
        \vspace{-0.2cm}
		\begin{tabular}{cccc}
			\toprule
			Method & Avg($R$) & Time/ms & Violation\\
			\midrule
			Proposed & -6.567 & 1.33 & 0 \\
			  No Correction & -9.325 & 1.24 & 12 \\ 
			\bottomrule
		\end{tabular}
        \vspace{-0.2cm} 
	\end{table} 
 
\subsubsection{Computational Efficiency Analysis}
We further verified the impact of the proposed AL scheme on training, with the results presented in Table \ref{table_7}. As indicated by the table, the proposed scheme achieves efficient training for complex system security margin estimation in significantly less time compared to traditional methods. Moreover, it ensures the performance of the proposed security margin estimator, reaffirming the applicability of the proposed AL framework in complex real power systems.

        \setlength{\tabcolsep}{3pt}	
        \vspace{-0.3cm}
        \begin{table}[htb]
        \small
		\caption{Efficiency Verification of Proposed AL-based Security Assessment and Margin Estimation}
		\label{table_7}
		\centering
        \vspace{-0.2cm}
		\begin{tabular}{ccccc}
			\toprule
			Train Method & Acc. & Spe. & $t_{\mathrm{total}}$/s & $t_{\mathrm{train}}$/s\\
			\midrule
			Proposed & 0.9365 &  0.9336 & 2931 & 3.232\\
			Normal & 0.9531 & 0.9090 & 8524 & 23.93\\ 
			\bottomrule
		\end{tabular}
        \vspace{-0.2cm}
	\end{table}

\section{Conclusion}\label{sec:5}
Motivated by the growing need for efficient and safe emergency control strategy formulation in complex power systems, this paper presents an innovative pre-formulation approach for short-term voltage stability control based on SRL. The proposed scheme comprises two key components: a security margin estimation module using NNs and a decision-making module based on SRL with gradient projection. The security margin estimator accurately characterizes the nonlinear boundary of feasible actions in power system emergency control scenarios, enhancing interpretability and providing a solid theoretical foundation for gradient-based corrections. Leveraging a dueling network architecture alongside active learning (AL) techniques significantly improves the scheme’s performance and practicality in complex systems. The gradient projection-based method cleverly integrates hard constraints associated with power system stability and data-driven reinforcement learning techniques, enhancing method security while ensuring efficiency. Empirical testing on the IEEE 39-bus system and GPG validates the excellent performance of the proposed approach. Future research will focus on the outlier events with low frequency and high risk in strategies training, and simplify the decision model to enhance the applicability of the proposed scheme in realistic power systems.
        
\bibliography{Ref.bib}{}
\bibliographystyle{ieeetr}
\end{document}